\documentstyle[amssymb,11pt]{article}

\def\theequation{\arabic{section}.\arabic{equation}}

\renewcommand{\theequation}{\thesection.\arabic{equation}}

\global\arraycolsep=1pt
\input epsf
\input{tcilatex}
\begin{document}

\font\cmss=cmss10 \font\cmsss=cmss10 at 7pt \hfill \hfill IFUP-TH/01-28


\vspace{10pt}

\begin{center}
{\Large {\bf A NOTE ON THE IMPROVEMENT AMBIGUITY\\[0pt]
OF THE STRESS TENSOR AND THE CRITICAL LIMITS\\[0pt]
\vspace{6pt} OF CORRELATION FUNCTIONS}} \vspace{10pt}

\bigskip \bigskip

{\sl D. Anselmi}

{\it Dipartimento di Fisica, Universit\`a di Pisa, via F. Buonarroti 2,
56126 Pisa, Italia}
\end{center}

\vskip 2truecm

\begin{center}
{\bf Abstract}
\end{center}

I study various properties of the critical limits of correlators containing
insertions of conserved and anomalous currents. In particular, I show that
the improvement term of the stress tensor can be fixed unambiguously,
studying the RG\ interpolation between the UV\ and IR\ limits. The removal
of the improvement ambiguity is encoded in a variational principle, which
makes use of sum rules for the trace anomalies $a$ and $a^{\prime }$. 
Compatible results follow from the analysis of the RG\
equations. I perform a number of self-consistency checks and discuss the
issues in a large set of theories.

\vspace{4pt}

\vskip 1truecm

\vfill\eject

\section{Introduction}

\setcounter{equation}{0}

In a large set of models, the renormalization-group (RG) flow interpolates
between well-defined ultraviolet (UV) and infrared (IR) fixed points, the
zeros of the beta function. The RG interpolation can be studied comparing
the UV and IR limits of a certain class of correlators. Finite operators
play a special role in this context, since they define central charges in
the conformal limits.

Among the finite operators, noticeable are the conserved currents, in
particular the stress tensor $T_{\mu \nu }$. When the theory contains scalar
fields $\varphi $, there exists an improvement operator 
\[
\Delta T_{\mu \nu }=(\partial _{\mu }\partial _{\nu }-\Box \delta _{\mu \nu
})\varphi ^{2}, 
\]
which mixes with $T_{\mu \nu }$ under renormalization. It is possible to
diagonalize this mixing \cite{browncollins,hathrell2}, and this makes the
improvement term finite as well. There exists a one-parameter family of
finite, conserved, spin-2, dimension-4 operators $T_{\mu \nu }(\eta )=T_{\mu
\nu }+\eta \Delta T_{\mu \nu }$. At the level of the Lorentz commutator
algebra, the operators $T_{\mu \nu }(\eta )$ are equivalent. At the level of
the correlation functions and operator-product expansions, they are not. For
example, in a conformal field theory, the embedding in external gravity is
fixed unambiguously by conformal invariance. This means that there is no
improvement arbitrariness in the UV and IR\ limits. In most models, the RG\
equations for $\eta $ extend the removal of the improvement arbitrariness
from the critical points to the intermediate energies.

A universal principle for the removal of the improvement ambiguity can be
formulated using the sum rules for the trace anomalies $a$ and $a^{\prime }$ 
\cite{234}. This is a sort of variational principle \cite{inv}, which fixes
a priviledged value $\bar{\eta}$ for $\eta $. We can distinguish two cases.

$i)$ When the improvement term survives in a critical limit (typically, the
UV), the value $\bar{\eta}$ determined by the variational principle
coincides with the value fixed by conformal invariance at criticality and
the RG\ equations. Matching the stress tensor at intermediate energies with
its UV limit removes the $\eta $-arbitrariness at all energies.

$ii)$ When the improvement term vanishes at the critical points, all
operators $T_{\mu \nu }(\eta )$ are in principle equally acceptable, but the
value $\bar{\eta}$ is still priviledged. Specifically, the minimum of $%
\Delta a^{\prime }(\bar{\eta})$ over the flow trajectories connecting the
same pair of fixed points is equal to $\Delta c$ in a class of models 
\footnote{%
This relation is empirically known to hold in massive gaussian models,
unitary and not unitary. Nevertheless, a satisfactory theoretical
understanding of this relation is still lacking.}.

There is a universal way to remove the $\eta $-arbitrariness and select a
unique stress tensor, in accord with all present knowledge.

In this paper I\ study this issue and other properties of the critical
limits of correlators. In section 2 I discuss the properties of the
improvement term and list the criteria for the removal of the $\eta $%
-ambiguity. In section 3 I illustrate the statements in a set of gaussian
models where calculations can be carreid over to the end. Then, I analyse
the RG equations in IR-free and UV-free theories. In all cases the parameter 
$\eta $ is fixed uniquely with the rules of section 2. In the appendix I
discuss other aspects of the critical limits of correlators containing
insertions of finite and non-finite operators. In particular, I show that
anomalous currents and the topological-charge density are finite in various
models.

\section{Removal of the $\eta$-ambiguity}

\setcounter{equation}{0}

The UV and IR limits of correlators containing insertions of the trace of
the stress-tensor have been studied systematically in ref. \cite{234}.
General sum rules for the central charges $a$ and $a^{\prime }$ have been
written. Particularly meaningful is the notion of flow invariant, that is to
say a flow integral, or combination of flow integrals, whose value depends
only on the extrema of the flow.

The theory is embedded in external gravity. The definition of $a$ and $%
a^{\prime }$ (and $c$) from the trace anomaly at criticality reads 
\begin{mathletters}
\[
\Theta _{*}={\frac{1}{(4\pi )^{2}}}\left[ c\,W^{2}-{\frac{a}{4}}\,{\rm G}+{%
\frac{2}{3}}\,a^{\prime }\,\Box R\right] , 
\]
where $W$ is the Weyl tensor and ${\rm G}=4R_{\mu \nu \rho \sigma }R^{\mu
\nu \rho \sigma }-16R_{\mu \nu }R^{\mu \nu }+4R^{2}$ is the Euler density.
The background metric is specialized to be conformally flat, i.e. $g_{\mu
\nu }=\delta _{\mu \nu }{\rm e}^{2\phi }$. The $\Theta $-correlators are
related to the $\phi $-derivatives of the induced action for the conformal
factor, which I denote with $\Gamma [\phi ]$. At criticality, $\Gamma [\phi
] $ depends only on the quantities $a$ and $a^{\prime }$. The sum rules for $%
\Delta a=a_{{\rm UV}}-a_{{\rm IR}}$ and $\Delta a^{\prime }=a_{{\rm UV}%
}^{\prime }-a_{{\rm IR}}^{\prime }$ can be written studying the critical
limits of the $\Theta $-correlators.

In this paper, I study specifically two sum rules for $\Delta a$, taken from
section 7 of \cite{234}. The first formula involves integrals of the two-
and three-point functions: 
\end{mathletters}
\begin{eqnarray}
\Delta a &=&{\frac{\pi ^{2}}{48}}\int {\rm d}^{4}x\,|x|^{4}\,\langle 
\widetilde{\Theta }(x)\,\widetilde{\Theta }(0)\rangle  \nonumber \\
&+&{\frac{\pi ^{2}}{48}}\int {\rm d}^{4}x\,{\rm d}^{4}y\,\,x^{2}\,y^{2}\,%
\left\{ \langle \widetilde{\Theta }(x)\,\widetilde{\Theta }(y)\,\widetilde{%
\Theta }(0)\rangle +\left\langle {\frac{\widetilde{\delta }\widetilde{\Theta 
}(x)}{\widetilde{\delta }\phi (y)}}\,\widetilde{\Theta }(0)\right\rangle
+2\left\langle {\frac{\widetilde{\delta }\widetilde{\Theta }(x)}{\widetilde{%
\delta }\phi (0)}}\,\widetilde{\Theta }(y)\right\rangle \right\} .
\label{sum1}
\end{eqnarray}
The second formula involves integrals of the two-, three- and four-point
functions: 
\begin{eqnarray}
\Delta a &=&{\frac{\pi ^{2}}{48}}\int {\rm d}^{4}x\,|x|^{4}\,\langle 
\widetilde{\Theta }(x)\,\widetilde{\Theta }(0)\rangle +{\frac{\pi ^{2}}{48}}%
\int {\rm d}^{4}x\,{\rm d}^{4}y\,{\rm d}^{4}z\,\left( x\cdot y\right) \left(
x\cdot z\right) \,\left\langle \widetilde{\Theta }(x)\,\widetilde{\Theta }%
(y)\,\widetilde{\Theta }(z)\,\widetilde{\Theta }(0)\right.  \nonumber \\
&+&2{\frac{{\widetilde{\delta }}\widetilde{\Theta }(x)}{{\widetilde{\delta }}%
\phi (y)}}\,\widetilde{\Theta }(z)\,\widetilde{\Theta }(0)+{\frac{{%
\widetilde{\delta }}\widetilde{\Theta }(x)}{{\widetilde{\delta }}\phi (0)}}\,%
\widetilde{\Theta }(y)\,\widetilde{\Theta }(z)+{\frac{{\widetilde{\delta }}%
\widetilde{\Theta }(y)}{{\widetilde{\delta }}\phi (z)}}\,\widetilde{\Theta }%
(x)\,\widetilde{\Theta }(0)+2{\frac{{\widetilde{\delta }}\widetilde{\Theta }%
(y)}{{\widetilde{\delta }}\phi (0)}}\,\widetilde{\Theta }(x)\,\widetilde{%
\Theta }(z)  \nonumber \\
&+&{\frac{\widetilde{\delta }^{2}\widetilde{\Theta }(x)}{\widetilde{\delta }%
\phi (y)\,\widetilde{\delta }\phi (z)}}\,\widetilde{\Theta }(0)+2{\frac{%
\widetilde{\delta }^{2}\widetilde{\Theta }(x)}{\widetilde{\delta }\phi (y)\,%
\widetilde{\delta }\phi (0)}}\,\widetilde{\Theta }(z)+{\frac{\widetilde{%
\delta }^{2}\widetilde{\Theta }(y)}{\widetilde{\delta }\phi (z)\,\widetilde{%
\delta }\phi (0)}}\,\widetilde{\Theta }(x)  \nonumber \\
&+&\left. 2{\frac{{\widetilde{\delta }}\widetilde{\Theta }(x)}{{\widetilde{%
\delta }}\phi (y)}}\,{\frac{{\widetilde{\delta }}\widetilde{\Theta }(z)}{{%
\widetilde{\delta }}\phi (0)}}+{\frac{{\widetilde{\delta }}\widetilde{\Theta 
}(x)}{{\widetilde{\delta }}\phi (0)}}\,{\frac{{\widetilde{\delta }}%
\widetilde{\Theta }(y)}{{\widetilde{\delta }}\phi (z)}}\right\rangle .
\label{sum2}
\end{eqnarray}

The notation is as follows. If $\varphi $ denotes generically the dynamical
fields of the theory, with conformal weight $h$, then the $\widetilde{\delta 
}/\widetilde{\delta }\phi $-derivatives are the derivatives with respect to $%
\phi $ at constant $\widetilde{\varphi }\equiv \varphi \,{\rm e}^{h\phi }$.
We have $\widetilde{\Theta }=-{\widetilde{\delta }S/\widetilde{\delta }\phi }
$, where $S$ denotes the action. It is understood that, after taking the $%
\phi $-derivatives of $\widetilde{\Theta }$, $\phi $ is set to zero.

I\ also study the $\Delta a^{\prime }$-sum rule

\begin{equation}
\Delta a^{\prime }={\frac{\pi ^{2}}{48}}\int {\rm d}^{4}x\,|x|^{4}\,\langle 
\widetilde{\Theta }(x)\,\widetilde{\Theta }(0)\rangle .  \label{deltaa'}
\end{equation}

The central charge $a$ in unambiguous at criticality, but $a^{\prime }$ is
ambiguous. This ambiguity disappears in the difference $\Delta a^{\prime }$,
which is a physical quantity. These facts have important implications in the
context of flow invariance and the dependence on the improvement ambiguity.

We can evaluate the above flow integrals using the one-parameter family of
stress tensors $T_{\mu \nu }(\eta )$. Two situations can occur.

If the improvement term of the stress tensor does not vanish at both
critical points, some non-trivial functions of $\eta $ are generated, which
depend also on the sum rule. I denote this dependence with a subscript $i$
and write $\Delta _{i}a(\eta )$. Formulas (\ref{sum1}) and (\ref{sum2})
define the functions $\Delta _{1}a(\eta )$ and $\Delta _{2}a(\eta )$,
respectively. Formula (\ref{deltaa'}) defines $\Delta a^{\prime }(\eta )$.
Since, however, $a$ is unambiguous at criticality, there must be a
priviledged value of $\eta $ which resolves the ambiguity and reproduces the
correct $\Delta a$. This value can be found studying the RG\ equations for
the parameter $\eta $, imposing conformal invariance at the critical points.

If the improvement term vanishes at both critical points, all values of $%
\eta $ are in principle acceptable. The functions $\Delta _{i}a(\eta )$ do
not depend on $i$ and $\eta $ and are identically equal to $\Delta a$.
Instead, since $a^{\prime }$ has no unambiguous definition at criticality,
the function $\Delta a^{\prime }(\eta )$ can depend on $\eta $. We know from
ref. \cite{inv} that the value $\bar{\eta}$ at which $\Delta a^{\prime
}(\eta )$ is minimum has particularly interesting properties. Using this, we
can remove the improvement ambiguity also in this case.

Both situations are resolved by a universal criterion for the removal of the
improvement ambiguity, encoded in a variational principle studied in \cite
{inv}, which expresses the independence of the flat-space theory from the
embedding in external gravity. When both this principle and the analysis of
the RG\ equations fix $\eta $, the results agree.

\medskip

{\bf Criterion for the removal of the improvement ambiguity.} Determine the
(unique) $\bar{\eta}$ which satisfies 
\begin{equation}
\left. {\frac{{\rm d}\Delta a^{\prime }(\eta )}{{\rm d}\eta }}\right| _{\eta
=\bar{\eta}}=0,\qquad \left. {\frac{{\rm d}\Delta _{i}a(\eta )}{{\rm d}\eta }%
}\right| _{\eta =\bar{\eta}}=0.  \label{varia}
\end{equation}
The functions $\Delta _{i}a(\eta )$ and $\Delta a^{\prime }(\eta )$ are at
most quadratic in $\eta $ (this will be shown explicitly in the next section%
\footnote{%
In even dimension greater than four, the $\eta $-polynomials can have a
higher degree. I\ am grateful to G. Festuccia for this remark.}), so the
condition (\ref{varia}) has one solution for every sum rule. The solution $%
\bar{\eta}$ does not depend on the sum rule. The correct stress tensor is $%
T_{\mu \nu }(\bar{\eta})$ and the correct value of $\Delta a$ is $\Delta
_{i}a(\bar{\eta})$, independently of $i$. This criterion fixes also $\Delta
a^{\prime }$ unambiguously.

\medskip

The integrals of (\ref{sum1}) and (\ref{sum2}) are assured to converge, when
there is no improvement ambiguity. When the stress tensor admits improvement
terms, instead, there can be a divergence in $\Delta a^{\prime }(\eta )$.
This divergence provides alternative criteria for the removal of the $\eta $%
-ambiguity (see below). If, on the other hand, the $\Theta $-correlators are
expanded perturbatively, the convergence of the term-by-term integration is
not assured. Observe that the resolution of the $\eta $-ambiguity is
intrinsically non-perturbative. A useful perturbative expansion can be
defined, although computations are not simple.

\medskip

{\bf Shortcuts and other criteria to remove the $\eta $-ambiguity.} The
value $\bar{\eta}$ does not depend on the sum rule and so it can be
determined from the simplest of those, i.e. $\Delta a^{\prime }(\eta )$. The 
$\Delta a$-sum rules involve more complicated flow integrals. In various
cases, $\bar{\eta}$ can be fixed by conformal invariance at the critical
points. In the nest sections I study the criteria for the removal of the
improvement ambiguity in a variety of models. These cover essentially all
cases. We can have the following behaviors:

$i$) the RG equations imply that the improvement term of $T_{\mu\nu}(\eta)$
survives at one of the critical points, where however the stress tensor is
uniquely fixed by conformal invariance;

$ii$) the RG equations imply that the improvement term of $T_{\mu\nu}(\eta)$
diverges at one of the critical points and the divergence disappears if $%
\eta $ is chosen appropriately;

$iii$) the improvement term vanishes at the critical points, but not
sufficiently quickly. The quantity $\Delta a^{\prime }$, should be finite,
because it is physically meaningful (although it is not a flow invariant 
\cite{inv}). The finiteness of $\Delta a^{\prime }(\eta )$ can fix $\eta $.
This can also be seen as a consequence of (\ref{varia}).

In all cases, the $\bar{\eta}$ fixed with the criteria ($i$), ($ii$) and ($%
iii$) coincides with the $\bar{\eta}$ of (\ref{varia}). In the next section
I present checks of this.

The forth situation is when the improvement term disappears quickly enough
at both critical points. When this happens, we have $\Delta _{i}a(\eta
)=\Delta a$ for every $i$ and $\eta $. The variational principle (\ref{varia}%
) applies also to this case, in the sense that it outlines a noticeable
value $\bar{\eta}$, such that $\Delta a^{\prime }(\bar{\eta})$ has the
properties studied in \cite{inv}. This behavior is studied in a
higher-derivative model.

\medskip

{\bf Modified, $\eta $-independent sum rules.} Following \cite{inv}, the
criterion (\ref{varia}) is equivalent to the $\eta $-independence of more
complicated sum rules. This illustrates that the removal of the $\bar{\eta}$%
-ambiguity fixed by (\ref{varia}) is compatible with the fact that the
quantum field theory in flat space is independent of the non-minimal
couplings to external gravity.

We proceed as follows. Using the fact that $\Delta _{i}a(\eta )$ is at most
quadratic in $\eta $, we write 
\begin{equation}
\Delta _{i}a(\eta )=\Delta _{i}a(\tilde{\eta})+(\eta -\tilde{\eta})\left. {%
\frac{{\rm d}\Delta _{i}a(\eta )}{{\rm d}\eta }}\right| _{\tilde{\eta}}+{%
\frac{1}{2}}(\eta -\tilde{\eta})^{2}\left. {\frac{{\rm d}^{2}\Delta
_{i}a(\eta )}{{\rm d}\eta ^{2}}}\right| _{\tilde{\eta}}.  \label{equa}
\end{equation}
The right-hand side is clearly independent of $\tilde{\eta}$.

Finding $\bar{\eta}$ according to (\ref{varia}), inserting it in (\ref{equa}%
) and renaming $\tilde{\eta}\rightarrow \eta $, we get 
\begin{equation}
\Delta a=\Delta _{i}a(\bar{\eta})=\Delta _{i}a(\eta )-{\frac{1}{2}}{\frac{%
\left( {\frac{{\rm d}\Delta _{i}a(\eta )}{{\rm d}\eta }}\right) ^{2}}{{\frac{%
{\rm d}^{2}\Delta _{i}a(\eta )}{{\rm d}\eta ^{2}}}}}.  \label{gensum}
\end{equation}
The final expression is an involved combination of flow integrals. It can be
seen as a generalized sum rule for $\Delta a$, in the spirit of the formulas
of \cite{inv}. The result is clearly independent of $\eta $ and gives the
correct value of $\Delta a$. In the generalized sum rule, we can chose for $%
\Delta _{i}a(\eta )$ any equivalent $\Delta a$-formula from ref. \cite{234};
for example, (\ref{sum1}) and (\ref{sum2}) of the present paper. The $i$%
-independence of the result can be rephrased in terms of equivalence
relations among the flow integrals. These involve correlators of $\Theta $
and the improvement operator.

\section{Checks and illustrative examples}

\setcounter{equation}{0}

In this section I study various examples, starting from simplest case,
namely the massive free scalar. A richer structure is exhibited by gaussian
non-unitary theories, where the issue of flow invariance is more apparent.
This model describes some qualitative features of physical theories with
several independent masses or dimensioned parameters. Then, I consider the $%
\varphi^4$-theory and asymptotically-free theories, supersymmetric and
non-supersymmetric. Finally, I comment on the most general case.

\medskip

{\bf Massive scalar field. }The action in external gravity is 
\[
S=\frac{1}{2}\int {\rm d}^{4}x~\sqrt{g}\left\{ g^{\mu \nu }\partial _{\mu
}\varphi ~\partial _{\nu }\varphi +\eta R\varphi ^{2}+m^{2}\varphi
^{2}\right\} . 
\]
Focusing on the conformal factor $\phi $ and eliminating a total derivative,
we can simplify the action and write 
\begin{equation}
S=\frac{1}{2}\int {\rm d}^{4}x\left\{ \left( \partial _{\mu }\widetilde{%
\varphi }\right) ^{2}+m^{2}\widetilde{\varphi }^{2}{\rm e}^{2\phi }+\left(
1-6\eta \right) \widetilde{\varphi }^{2}\left( \Box \phi +(\partial _{\mu
}\phi )^{2}\right) \right\} ,  \label{acca}
\end{equation}
where $\widetilde{\varphi }=\varphi \,{\rm e}^{\phi }$. We need 
\begin{equation}
\widetilde{\Theta }=-\frac{\widetilde{\delta }S}{\widetilde{\delta }\phi }%
=-m^{2}\widetilde{\varphi }^{2}{\rm e}^{2\phi }+\frac{1}{2}\left( 1-6\eta
\right) \left[ \Box \left( \widetilde{\varphi }^{2}\right) -2\partial _{\mu
}\left( \widetilde{\varphi }^{2}\partial _{\mu }\phi \right) \right] ,
\label{scalar}
\end{equation}
where $\widetilde{\delta }$ is the $\phi $-derivative at fixed $\widetilde{%
\varphi }$ (check \cite{234} for definitions), and the first two derivatives
of $\widetilde{\Theta }$ with respect to $\phi $.

The calculations give 
\[
\Delta _{1}a(\eta )=-{\frac{89}{360}+3\eta }-9\eta
^{2},~~~~~~~~~~~~~~~~~~~\Delta _{2}a(\eta )=-{\frac{37}{180}+}\frac{5}{2}%
\eta -{\frac{15}{2}}\eta ^{2}. 
\]
The condition (\ref{varia}) gives the (expected) value $\bar{\eta}=1/6$ in
both cases and $\Delta _{1}a(\bar{\eta})=\Delta _{2}a(\bar{\eta}%
)=1/360=\Delta a$. It is well-known that the value $\bar{\eta}=1/6$ is such
that the action (\ref{acca}) is conformal at $m=0$. The correct stress
tensor can be fixed, more simply, by requiring that $\widetilde{\Theta}$ be
zero in the UV limit. This is a check that the criterion (\ref{varia}) gives
the same result as conformal invariance at the critical points, expressed by
shortcut ($i$). On the other hand, $\Delta _{1}a(\eta )$ and $\Delta
_{2}a(\eta )$ do not coincide for $\eta \neq \bar{\eta}$.

The coincidences of the values of $\bar{\eta}$ determined by $\Delta
_{1,2}a(\eta )$ and the equality of $\Delta _{1,2}a(\bar{\eta})$ are
non-trivial. They are due to identities among the flow integrals. An
illustrative example is 
\[
m^{2}\int {\rm d}^{4}x\,{\rm d}^{4}y\,\,x^{2}\langle \varphi
^{2}(x)\,\varphi ^{2}(y)\,\varphi ^{2}(0)\rangle =2\int {\rm d}%
^{4}x\,\,x^{2}\langle \varphi ^{2}(x)\,\varphi ^{2}(0)\rangle , 
\]
which is relevant for the calculation of $\Delta _{1}a(\eta )$. This
identity can be verified directly or seen as a consequence of dimensional
counting (each integral has the form const.$/m^{2}$), combined with the
property that an insertion of $\int {\rm d}^{4}x\,m^{2}\varphi ^{2}(x)$ is
equivalent to the derivative $-m\partial /\partial m$. A similar
cancellation takes place in $\Delta _{2}a(\eta )$. The variational principle
(\ref{varia}) ``knows'' about such relations.

Let us now study $\Delta a^{\prime }$. The explicit calculation shows that a
coefficient is infinite. Precisely: 
\[
\Delta a^{\prime }(\eta )={-}\frac{3}{40}+\frac{1}{2}\eta +\left( 1-6\eta
\right) ^{2}\infty . 
\]
In the $\Delta a$-sum rule (\ref{sum1}), the infinite term is compensated by
a contribution coming from the flow integral of $\langle {\widetilde{\delta }%
\widetilde{\Theta }/\widetilde{\delta }\phi }~\Theta \rangle $ and the sum
is finite for each value of $\eta $. An analogous compensation occurs in (%
\ref{sum2}). Observe that (\ref{varia}), applied to $\Delta a^{\prime }$,
still fixes $\bar{\eta}$ to $1/6$, so that, correctly, $\Delta a^{\prime }(%
\bar{\eta})=1/120$ \cite{cea,inv}.

The quantity $\Delta a^{\prime }$ is much less restricted than $\Delta a$.
For example, it can depend on the flow connecting the two fixed points \cite
{inv}. Still, it is a physically meaningful quantity and characterizes the
flow. The infinity of $\Delta a^{\prime }(\eta )$ is not a ``divergence'' to
be removed. The correct value of $\Delta a^{\prime }$ must be finite. In the
theory at hand (but also in the $\varphi ^{4}$-theory and other models
discussed below), finiteness of $\Delta a^{\prime }$ fixes $\bar{\eta}$. We
have a check that the $\bar{\eta}$s fixed with shortcut ($iii$) and any of
the (\ref{varia}) coincide.

\medskip

{\bf Flow invariance.} Examples of flows interpolating between the same
fixed points are easy to construct. An as illustration, take non-Abelian
Yang-Mills theory with group $G=SU(N_{c})$, $N_{f}$ massless quarks and $%
M_{f}$ massive quarks in the fundamental representation. In the large $N_{c}$
limit, with $N_{f}/N_{c}\lesssim 11/2$ fixed, the theory is UV-free and has
an interacting IR fixed point. Indeed, at low energies, the massive fermions
decouple and the beta function 
\[
\beta =-{\frac{1}{6\pi }}(11N_{c}-2N_{f})\,\alpha ^{2}+{\frac{25}{(4\pi )^{2}%
}}N_{c}^{2}\,\alpha ^{3}+{\cal O}(\alpha ^{4}) 
\]
has a second zero. The higher-loop terms can be neglected in the given large-%
$N_{c}$ limit.

The UV and IR fixed points do not depend on the values of the masses of the $%
M_f$ massive quarks. For each set of values of the masses we have a
different flow interpolating between the same conformal field theories.

At the computational level, it is not easy to study the sum rules (\ref{sum1}%
) and (\ref{sum2}) in this model. A treatable perturbative expansion of the
flow integrals of (\ref{sum1}) and (\ref{sum2}) has still to be developed.
Gaussian higher-derivative theories, on the other hand, provide an
interesting laboratory of flows interpolating between the same fixed points.
Calculations are still lengthy, but doable.

\medskip

{\bf Higher-derivative scalar field.} The lagrangian of the theory is 
\begin{equation}
{\cal L}={\frac{1}{2}}\left[(\Box\varphi)^2+\beta
m^2(\partial_{\mu}\varphi)^2 +m^4\varphi^2\right].  \label{higher}
\end{equation}
The embedding in external gravity gives 
\begin{equation}
{\cal L}={\frac{1}{2}}\sqrt{g}\left(\varphi\Delta_4\varphi+ \beta
m^2(\partial_{\mu}\varphi)(\partial_{\nu}\varphi)g^{\mu\nu} +\eta R m^2
\varphi^2+m^4 \varphi^2\right),  \label{uno}
\end{equation}
where the differential operator 
\begin{equation}
\Delta_4=\nabla^2\nabla^2+2\nabla_{\mu}\left[ R^{\mu\nu}-{\frac{1}{3}}%
g^{\mu\nu}R\right] \nabla_{\nu}  \label{due}
\end{equation}
is such that $\sqrt{g}\Delta_4$ is conformally invariant (see for example 
\cite{riegert}) .

I do not consider non-minimal couplings of the form $R^2\varphi^2$. Their
coefficients can be set to zero imposing $\Theta=0$ at criticality, as in
the previous example. The non-minimal coupling $m^2R\varphi^2$, instead,
disappears both in the UV limit ($m\rightarrow 0$) and IR limit ($%
m\rightarrow \infty$ and $\varphi\rightarrow 0$, keeping the mass term $%
m^4\varphi^2$ bounded).

I perform two calculations, with (\ref{sum1}) and (\ref{sum2}). The relevant
operator is 
\[
\overline{\Theta }=-{\frac{\delta S}{\delta \phi }}=-\beta m^{2}(\partial
_{\mu }\varphi )^{2}{\rm e}^{2\phi }-2m^{4}\varphi ^{2}{\rm e}^{4\phi
}+3\eta m^{2}\left[ {\rm e}^{\phi }\Box \left( \varphi ^{2}{\rm e}^{\phi
}\right) +{\rm e}^{\phi }\varphi ^{2}\Box {\rm e}^{\phi }\right] . 
\]
Tilded quantities are equal to untilded quantities in this model, since the
canonical weight of the higher-derivative scalar field is zero.

Cubic and quartic terms in $\eta$ do not contribute to (\ref{sum1}) and (\ref
{sum2}). The improvement term (in $\widetilde{\Theta}$ and it its $\phi$%
-derivatives) carries a $\Box$. Using integrations by parts, the boxes can
be moved and act on the degree-4 polynomials $x^2y^2$ or $(x\cdot y)(x\cdot
z)$. Three boxes kill the polynomials and therefore the integral. This
observation implies that the condition (\ref{varia}) always has a unique
solution.

The sum rules (\ref{sum1}) and (\ref{sum2}) give 
\[
\Delta _{1}a(\eta )=\Delta _{2}a(\eta )=-{\frac{7}{90}}=\Delta a, 
\]
independently of $\eta $. I recall that in this model, $a_{{\rm UV}}=-7/90$ 
\cite{antonia} and $a_{{\rm IR}}=0$.

\medskip

The calculations, lengthy and cumbersome, have been done with Mathematica. I
do not report here intermediate results, because they do not seem to be
particularly instructive.

The flow invariance of $\Delta _{i}a(\eta )$ and the cancellation of the $%
{\cal O}(\eta )$-terms and ${\cal O}(\eta ^{2})$-terms in (\ref{sum1}) and (%
\ref{sum2}) are consequences of non-trivial identities among flow integrals.
Each term of (\ref{sum1}) and (\ref{sum2}) separately violates these
properties. As for the quantity $\Delta a^{\prime }$, we have 
\begin{eqnarray*}
\Delta a^{\prime }(\eta ) &=&{\frac{%
1+17r^{2}-17r^{4}-r^{6}+10(1+r^{2}+r^{4}+r^{6})\ln r}{40(r^{2}-1)^{3}}}+\eta
U(r)+\eta ^{2}V(r), \\
U(r) &=&-3r{\frac{1-r^{4}+2(1+r^{4})\ln r}{2(r^{2}-1)^{3}}},\qquad
V(r)=9r^{2}{\frac{1-r^{2}+(1+r^{2})\ln r}{(r^{2}-1)^{3}}},
\end{eqnarray*}
where $r$ is defined by $\beta =r+1/r$. $r$ is the unique dimensionless
parameter of the theory, besides the improvement coefficient $\eta $. Since $%
\Delta a^{\prime }(\eta )$ is finite for every $\eta $, none of the
shortcuts of the previous section applies. All values of $\eta $ are in
principle acceptable, but the value of $\eta $ which minimizes $\Delta
a^{\prime }(\eta )$ is priviledged, in the sense that it has various
interesting properties, outlined in \cite{inv}. We have 
\[
\bar{\eta}(r)=-\frac{U(r)}{2V(r)}={\frac{1-r^{4}+2(1+r^{4})\ln r}{12r\left(
1-r^{2}+(1+r^{2})\ln r\right) }} 
\]
and \cite{inv} 
\[
\Delta a^{\prime }(\bar{\eta})=-{\frac{%
(r^{2}-1)^{2}(3r^{4}-26r^{2}+3)+(r^{8}+18r^{6}-18r^{2}-1)\ln
r^{2}-10r^{2}(r^{4}+1)\ln ^{2}r^{2}}{40(r^{2}-1)^{3}(r^{2}\ln r^{2}+\ln
r^{2}-2r^{2}+2)}.} 
\]
The value $\Delta a^{\prime }(\bar{\eta})$ depends on $r$, which means that
it not a flow invariant. Its minimum coincides with $\Delta c=-1/15$ \cite
{inv}.

In conclusion, the ``good'' stress tensor of the theory (\ref{higher}) is 
\begin{eqnarray*}
T_{\mu \nu } &=&-\partial _{\nu }\Box \varphi \,\partial _{\mu }\varphi
-\partial _{\mu }\Box \varphi \,\partial _{\nu }\varphi +2\Box \varphi
\,\partial _{\mu }\partial _{\nu }\varphi +\frac{2}{3}\partial _{\mu
}\partial _{\nu }\partial _{\alpha }\varphi \,\partial _{\alpha }\varphi -{%
\frac{4}{3}}\partial _{\mu }\partial _{\alpha }\varphi \,\partial _{\nu
}\partial _{\alpha }\varphi \\
&&+\delta _{\mu \nu }\left[ \frac{1}{3}\partial _{\alpha }\Box \varphi
\,\partial _{\alpha }\varphi +\frac{1}{3}\left( \partial _{\alpha }\partial
_{\beta }\varphi \right) ^{2}-\frac{1}{2}\left( \Box \varphi \right) ^{2}-{%
\frac{m^{4}}{2}}\varphi ^{2}\right] +\beta m^{2}\left( \partial _{\mu
}\varphi \,\partial _{\nu }\varphi -{\frac{\delta _{\mu \nu }}{2}}(\partial
_{\alpha }\varphi )^{2}\right) \\
&&-\bar{\eta}(r)m^{2}\left( \partial _{\mu }\partial _{\nu }-\Box \delta
_{\mu \nu }\right) \varphi ^{2}.
\end{eqnarray*}
and does not contain any more parameters than the flat-space action (\ref
{higher}).

\medskip {\bf The $\varphi^4$-theory.} The renormalization mixing between
the stress tensor and its improvement term in the $\varphi^4$-theory has
been studied in detail by Brown and Collins \cite{browncollins} and Hathrell 
\cite{hathrell2}. In the formulas below, the dimensional regularization
technique and the minimal subtraction scheme are understood.

The parameter $\eta $ satisfies the inhomogeneous RG equation \cite
{hathrell2} 
\begin{equation}
\mu {\frac{{\rm d}\eta }{{\rm d}\mu }}-\delta (\lambda )\eta =\beta _{\eta
}(\lambda )\equiv -\delta (\lambda )d(\lambda ).  \label{rg}
\end{equation}
Here $\delta (\lambda )$ is the anomalous dimension of the composite
operator $\varphi ^{2}$, while $d(\lambda )$ is determined by the simple
pole in the $\varphi ^{4}$-$\Box \varphi ^{2}$ renormalization mixing.
Precisely, 
\[
{\frac{\mu ^{4-n}[\varphi ^{4}]}{4!}}={\frac{(n-4)}{\hat{\beta}}}\left\{ {%
\frac{\lambda _{0}\varphi _{0}^{4}}{4!}}-{\frac{\gamma }{n-4}}[{\rm E}]-{%
\frac{d+L_{d}}{n-4}}\Box [\varphi ^{2}]\right\} , 
\]
where $n$ is the space-time dimension, [E] is the $\varphi $-field equation, 
$\gamma $ is the $\varphi $-anomalous dimension, $\hat{\beta}=(n-4)\lambda
+\beta (\lambda )$, $\beta (\lambda )$ is the beta function, and $L_{d}$
denotes the poles higher that the simple one. The subscript 0 denotes bare
quantities, and the square brakets denote renormalized operators. The trace
of the stress tensor reads in four dimensions 
\[
\widetilde{\Theta }=-\beta {\frac{[\varphi ^{4}]}{4!}}-\gamma [{\rm E}%
]+(\eta -d)\Box [\varphi ^{2}]. 
\]
The equation (\ref{rg}) can be decomposed in the following way: 
\[
\eta =\widetilde{\eta }(\lambda )+\eta ^{\prime }v(\lambda ), 
\]
where $\eta ^{\prime }$ is finite ($\mu {\rm d}\eta ^{\prime }/{\rm d}\mu =0$%
), $\widetilde{\eta }$ is a particular solution of (\ref{rg}), fixed
conventionally so that $\widetilde{\eta }(0)=0$, and $v$ satisfies the
homogeneous equation: 
\begin{equation}
v(\lambda )=\exp \left( \int^{\lambda }{\frac{\delta (\lambda ^{\prime })}{%
\beta (\lambda ^{\prime })}}{\rm d}\lambda ^{\prime }\right) .
\label{vlambda}
\end{equation}
It is not necessary to specify the second extremum of integration, which can
be absorbed in the factor $\eta ^{\prime }$. The function $v(\lambda )$ is
related to the renormalization constant of the operator $\varphi ^{2}$.

The surviving finite constant $\eta^{\prime}$ parametrizes the stress-tensor
ambiguity, which reads 
\begin{equation}
T_{\mu\nu}(\eta^{\prime})=T_{\mu\nu}(0)- {\frac{1}{3}}\eta^{\prime}v(%
\lambda)(\partial_{\mu}\partial_{\nu} -\delta_{\mu\nu}\Box)[\varphi^2].
\label{etapa}
\end{equation}
It follows immediately from (\ref{vlambda}) that if $T_{\mu\nu}(0)$ is
finite ($\mu\, {\rm d}T_{\mu\nu}(0)/{\rm d}\mu=0$), then $%
T_{\mu\nu}(\eta^{\prime})$ is also finite.

In \cite{browncollins} it was observed that $\eta ^{\prime }$ can be
consistently set to zero. In \cite{hathrell2} it was remarked that $\eta
^{\prime }$ should be fixed ``by experiment'', since it is the coefficient
of the non-minimal coupling to external gravity. Here we want to see if
there is a reason why $\eta ^{\prime }$ should be set a priori to a
particular value.

There is strong evidence that the $\varphi^4$-theory is non-perturbatively
trivial. Even if we cannot view this theory as an RG interpolation between a
UV and a IR fixed point, we can make a couple of general observations, which
apply also to more general cases studied below.

To the lowest order, we have \cite{hathrell2} 
\[
\delta (\lambda )={\frac{\lambda }{(4\pi )^{2}}}+{\cal O}(\lambda
^{2}),~~~~~~~~~~\beta (\lambda )=3{\frac{\lambda ^{2}}{(4\pi )^{2}}}+{\cal O}%
(\lambda ^{3}),~~~~~~~~~~\beta _{\eta }(\lambda )=-{\frac{1}{36}}{\frac{%
\lambda ^{4}}{(4\pi )^{8}}}, 
\]
so that 
\[
\eta =-{\frac{1}{288}}{\frac{\lambda ^{3}}{(4\pi )^{6}}}+{\cal O}(\lambda
^{4})+\eta ^{\prime }\lambda ^{1/3}\left( 1+{\cal O}(\lambda )\right) . 
\]
Let us consider the flow integral (\ref{deltaa'}), which defines $\Delta
a^{\prime }(\eta )$. In the absence of information about the UV, we can
study the convergence of this integral around the IR limit. Using the
perturbative values given above and the Callan-Symanzik equations for the
pair of operators $(\varphi ^{4},\Box \varphi ^{2})$, the behavior of the
integral around the IR is 
\[
\Delta a^{\prime }(\eta )\sim \int^{\infty }{\rm d}t\left( {\frac{a_{1}}{%
t^{4}}}+\eta ^{\prime }{\frac{a_{2}}{t^{10/3}}}+\eta ^{\prime 2}{\frac{a_{3}%
}{t^{2/3}}}\right) , 
\]
where the $a_{i}$ are numerical factors. We see that the ${\cal O}(\eta
^{\prime 2})$-contribution diverges in the IR extremum of integration. Since
the integrand is non-negative, this divergence cannot be cured by
contributions from intermediate energies or by a hypothetical second fixed
point (which exists in the models studied below, to which similar
considerations apply). Therefore, the only value compatible with a finite $%
\Delta a^{\prime }$ is $\eta ^{\prime }=0$.

This case is different from the case of a free-massive scalar field. Here
the improvement term of (\ref{etapa}) does disappear at criticality (at a
velocity $\lambda (t)^{1/3}$, where $\lambda (t)\sim 1/t$, $t=\ln |x|\mu $),
but it does not disappear sufficiently quickly for the sum rule to converge.
This forces $\eta ^{\prime }$ to be zero, by shortcut ($iii$).

\medskip

The $\varphi ^{4}$-interaction can be non-trivial in several models, which
may admit conformal windows. In particular, interesting cases are the
supersymmetric models, with or without superpotential. Supersymmetry is not
necessary to the logic of the arguments below, but it simplifies the
examples.

\medskip{\bf N=1 supersymmetric QCD.} I consider now N=1 supersymmetric QCD
with gauge group $G=SU(N_c)$ and $N_f$ quark and antiquark superfields in
the fundamental representation. The theory has no superpotential and a
unique coupling constant $g$. For $N_f<3N_c$ the theory is asymptotically
free. The mass operator $\overline{\varphi}\varphi$, which is essential for
the improvement term, is the lowest component of the Konishi superfield \cite
{andrei}. Since the axial currents have no anomalous dimension at the
one-loop order (see the appendix), this is true also of $\overline{\varphi}%
\varphi$. The two-loop contribution to the anomalous dimension $\delta$ can
be found in \cite{martin}: 
\[
\delta(g)=4\left(N_c^2-1\right) N_f\left({\frac{g^2}{16\pi^2}}\right)^2+%
{\cal O}(g^6). 
\]
The structure of the RG equation for $\eta$ and the $\eta^{\prime}$%
-ambiguity of the stress tensor are the same as in (\ref{rg}) and (\ref
{etapa}). In particular, the function $v$ is given by the analogue of (\ref
{vlambda}). The one-loop beta function is $\beta=-g^3\left(3N_c-N_f%
\right)/(16\pi^2)+{\cal O}(g^5)$, so that the function 
\begin{equation}
v(g)=\exp\left(-{\frac{g^2}{8\pi^2}}{\frac{N_f(N_c^2-1)}{(3N_c-N_f)}} +{\cal %
O}(g^4)\right)  \label{tc1}
\end{equation}
tends to unity at $g\rightarrow 0$. Instead, $\beta_{\eta}(g)$ goes to zero
at least as fast as $g^8$. We conclude that $\eta\rightarrow \eta^{\prime}$
in the UV limit, so that the improvement term of (\ref{etapa}) survives at
criticality. As in the case of the free massive scalar field, this forces $%
\eta^{\prime}$ to be set to zero, by shortcut ($i$).

\medskip

{\bf Supersymmetric theories with a superpotential.} The superpotential
gives a one-loop contribution to the anomalous dimension of the Konishi
operator. An example of UV-free supersymmetric theory with superpotential
and a well-defined IR fixed point is the theory obtained adding mesonic
fields $M_i^j$ to the N=1 supersymmetric QCD. The meson superfields interact
with the quarks $q_i$ and $\bar{q}^j$ by means of a superpotential $fM^i_jq_i%
\bar{q}^j$ \cite{kogan,universal}. In complete generality, denoting the
superpotential couplings by $Y_{ijk}$, the one-loop anomalous dimension of
the mass operator $\bar{\varphi}\varphi$ is 
\begin{equation}
\delta(Y)={\frac{3}{16\pi^2}}|Y|^2,  \label{deltay}
\end{equation}
$|Y|$ being defined by $Y_{ijk}Y^{ijl}=|Y|^2\delta_k^l$. Solving the RG
equations around the UV fixed point \cite{kogan} and applying (\ref{vlambda}%
), we have 
\begin{equation}
v\sim |t|^c,  \label{tc}
\end{equation}
for $t\rightarrow -\infty$, with $c$ positive numerical constant. The
situation is even worse than in the previous model, where the superpotential
was absent: the improvement term of the stress tensor diverges in the
free-field limit. This forces again to set $\eta^{\prime}=0$, by shortcut ($%
ii$).

\medskip

{\bf Asymptotically-free theories and flows with interacting UV fixed points.%
} The arguments of the previous two cases apply to the most general
asymptotically-free theory with scalar fields, supersymmetric or not. The
anomalous dimension of the improvement term can have a vanishing one-loop
contribution or a non-vanishing one-loop contribution. In either case, its
first radiative correction is positive. On the other hand, the first term of
the beta function is negative. Then, $v(t)$ behaves as in (\ref{tc1}) or (%
\ref{tc}). In the free-field limit, the improvement term of the stress
tensor is finite and non-vanishing or divergent. This fixes $\eta ^{\prime }$%
.

The same can be said of flows with interacting UV fixed points, where $%
\delta _{{\rm UV}}$ can be non-vanishing. We can conclude, in full
generality, that in unitary models the UV behavior of $T_{\mu \nu }(\eta
^{\prime })$ is unacceptable, unless the $\eta ^{\prime }$-term of (\ref
{etapa}) is suitably fixed, according to the rules of sect. 2.

\section{Conclusions}

A proper RG interpolation between the UV and IR fixed points removes the
improvement ambiguity of the stress tensor. The general criterion for this
removal is encoded in the variational principle (\ref{varia}). In various
cases suitable shortcuts can be more efficient. I have analysed concrete
examples, one for each relevant situation. In particular, in
asymptotically-free theories, the RG equations imply that the improvement
term survives at one critical point or diverges there, unless the
improvement parameter is suitably fixed. In IR free theories, the
improvement term does disappear at criticality, but not sufficiently
quickly. The behavior of gaussian higher-derivative theories shows that
there are cases in which all improved stress tensors are in principle
acceptable. Nevertheless, the criterion (\ref{varia}) outlines a priviledged
stress tensor also in this case. In conclusion, we can always consistently
remove the improvement ambiguity with the rules of section 2.

\section{Appendix. Other remarks about the critical limits of correlators}

In this appendix I\ consider other ambiguities in the critical limits of
correlators. Let us assume that two observers study the same model using
different renormalization schemes. We want to know what amount of
information the observers can objectively compare and how many quantities
they need to normalize before the comparison. I\ consider correlators of
composite operators and\ distinguish the cases of finite and non-finite
operators.

Let ${\cal O}$ be a multiplicatively renormalized operator. The
Callan-Symanzik equations imply that the two-point function can be written
in the form 
\begin{equation}
\langle {\cal O}(x)\,{\cal O}(0)\rangle ={\frac{1}{|x|^{2d}}}Z^{2}\left(
\alpha (1/|x|),\alpha (\mu )\right) G(\alpha (1/|x|)),  \label{twopt}
\end{equation}
where $d$ is the canonical dimension of ${\cal O}$, $Z$ is the
renormalization constant and $\alpha (\lambda )$ is the running coupling
constant at the energy scale $\lambda $. Now, in the UV (respectively, IR)
limit, namely $|x|\rightarrow 0$ ($|x|\rightarrow \infty $), $\alpha (1/|x|)$
tends to the critical value $\alpha _{{\rm UV\,(IR)}}$. If ${\cal O}$ is not
finite (i.e. $Z\neq 1$), then the limit depends on $\alpha (\mu )$ and the
subtraction scheme. The values $\alpha _{{\rm UV\,(IR)}}$ are themselves
scheme dependent.

Among the finite operators, we distinguish conserved currents, anomalous
(classically conserved) currents and others. Suppressing the space-time
indices, if ${\cal O}$ is a conserved current, the UV and IR limits of (\ref
{twopt}) have the structure 
\begin{equation}
\langle {\cal O}(x)\,{\cal O}(0)\rangle _{{\rm UV\,(IR)}}\sim {\frac{%
G(\alpha _{{\rm UV\,(IR)}})}{(x^{2})^{d}}}.  \label{oo}
\end{equation}
The quantities $G(\alpha _{{\rm UV\,(IR)}})$ (called primary central charges 
\cite{central}) carry information about the conformal fixed points. The
scheme dependence of $\alpha _{{\rm UV\,(IR)}}$ is compensated by an equal
and opposite scheme dependence of $G$, so that $G(\alpha _{{\rm UV\,(IR)}})$
is scheme independent. Similar considerations extend to correlators with
more insertions. When the two observers compare their results, they have to
find the same answer.

Classically-conserved anomalous currents\ can be finite. In an
asymptotically-free theory, for example, the renormalization constant $Z_{5}$
of the axial current $J_{5}^{\mu }$ resums non-perturbatively to a finite
function $C(\alpha )$. Finiteness can be formally recovered multiplying $%
J_{5}^{\mu }$ by $C^{-1}(\alpha )$. Then, in (\ref{twopt}) the
renormalization constant can be non-perturbatively replaced by unity. Scalar
operators can be finite as well. An example is the topological-charge
density. Details are given below. The function $G(\alpha (1/|x|))$ tends to
a constant in the free-field limit and behaves like a power of $x^{2}\mu ^{2}
$ around the interacting critical limit. Formula (\ref{oo}) is upgraded to
the more general expression 
\begin{equation}
\langle {\cal O}(x)\,{\cal O}(0)\rangle _{{\rm UV\,(IR)}}\sim {\frac{G_{{\rm %
UV\,(IR)}}}{(x^{2})^{d}(x^{2}\mu ^{2})^{h_{{\rm UV\,(IR)}}}}}.  \label{oo2}
\end{equation}
Here the critical limits are unambiguous once the scale $\mu $ is normalized
($\mu $ plays the role of the RG invariant scale, e.g. $\Lambda _{{\rm QCD}}$%
). Two observers can compare their results, once they agree on the
definition of the reference scale. It is possible to define ``secondary''
central charges \cite{central}, where the $\mu $-normalization is simplified
away.

Finally, the critical limits of correlators containing insertions of
non-finite operators provide one piece of information less \cite{central},
since a non-finite operator needs to be normalized at some reference energy.
In (\ref{twopt}) this is emphasized by the $\alpha (\mu )$-dependence
surviving in the limits $|x|\rightarrow 0$ and $|x|\rightarrow \infty $. 

These observations apply to operators whose 
correlators have power-behaved 
critical limits. 
Logarithmic behaviors are not unfrequent, however.
The improvement term of the stress tensor often exhibits a logarithmic 
behavior: check the
$\bar{\varphi}\varphi$-two-point function in $a$)
the $\varphi^4$-theory around the IR and $b$) supersymmetric theories 
with superpotential around the UV (see (3.11)).

\medskip

{\bf Anomalous currents.} Anomalous currents can be finite operators, and
therefore have unambiguous critical limits, of the form (\ref{oo2}). This
paragraph extends a discussion of Collins \cite{collins} to singlet currents
and the topological-charge density.

I consider the axial current in an asymptotically-free gauge theory. I
assume that the current is conserved at the classical level. The inclusion
of mass terms is straightforward. The anomaly equation 
\begin{equation}
\partial_{\mu} J_5^{\mu} -{\frac{g^2N_f}{16\pi^2}}F\widetilde{F}=\bar{\psi}%
\gamma_5 {\frac{\delta_l S}{\delta\bar{\psi}}}+{\frac{\delta_r S}{\delta \psi%
}} \gamma_5\psi={\rm finite}  \label{anomaly}
\end{equation}
and the definition $[J_5^{\mu}]=Z_5 J_5^{\mu}$ imply the relations 
\[
[\partial_{\mu} J_5^{\mu}]=Z_5 \partial_{\mu} J_5^{\mu},~~~~~~~~~~~~~ {\frac{%
g^2N_f}{16\pi^2}}[F\widetilde{F}]=\left(Z_5-1\right) \partial_{\mu}
J_5^{\mu} +{\frac{g^2_0N_f}{16\pi^2}}F\widetilde{F} 
\]
Calling ${\cal O}_1=\partial_{\mu} J^{\mu}_5$ and ${\cal O}_2= g^2F%
\widetilde{F}/(16\pi^2)$, we have 
\[
[{\cal O}_i]=Z_{ij}{\cal O}_j,~~~~~~~~Z_{ij}=\left(\matrix{Z_5 & 0 \cr
Z_5-1 & 1}\right). 
\]
Consider the two-point function $\langle
[J_5]^{\mu}(x)\,[J_5]^{\nu}(0)\rangle $. At the one-loop order it has a
conformal-invariant form, namely 
\begin{equation}
\langle [J_5^{\mu}](x)\,[J_5^{\nu}](0)\rangle =A(g^2){\frac{\delta^{\mu\nu}-
2{x^{\mu} x^{\nu}/ x^2}}{(x^2)^{3+\delta_5(g^2)}}}+{\cal O}(g^4).
\label{two}
\end{equation}
The one-loop conformal invariance is assured by the Callan-Symanzik
equations. Indeed, the conformal-violating term in 
\[
\mu{\frac{\partial}{\partial \mu}}+\beta(g){\frac{\partial}{\partial g}}
+2\delta_5(g^2) 
\]
is $\beta\,\partial/\partial g$. Since $\beta={\cal O}(g^3)$, this term is
irrelevant at the one-loop order.

Taking two divergences of (\ref{two}), using the anomaly equation (\ref
{anomaly}) and excluding the coincident point, we get 
\[
\left\langle [\partial_{\mu} J^{\mu}_5](x)\,\,[\partial_{\nu} J^{\nu}_5](0)
\right\rangle = {\frac{g^4N_f^2}{(4\pi)^4}} \left\langle [F\widetilde{F}%
](x)\,\, [F\widetilde{F}](0)\right\rangle=-4A(g^2){\frac{ \delta_5(g^2)
\left(2+\delta_5(g^2)\right) }{(x^2)^{4+\delta_5(g^2)}}}. 
\]
Since $A(g^2)={\cal O}(1)$, we conclude $\delta_5(g^2)={\cal O}(g^4)$. This
result is unaffected by the presence of masses or other super-renormalizable
parameters, but does not hold when the conservation of $J_5^{\mu}$ is
violated at the classical level by marginal operators, such as in
supersymmetric theories with a superpotential: see (\ref{deltay}).

Now, we observe that the renormalization constant $Z_5$ has a finite limit
when the cut-off is sent to infinity. We can see this using the
dimensional-regularization technique, but it is more explicit to write the
limit in the familiar cut-off notation. Precisely, 
\[
\lim_{\Lambda\rightarrow\infty}
Z_5(g(\Lambda),g(\mu))=\lim_{\Lambda\rightarrow\infty}
\exp\left(-\int^{g(\Lambda)}_{g(\mu)}{\frac{\delta_5(g^{\prime})}{%
\beta(g^{\prime})}}{\rm d} g^{\prime}\right)=C(g^2)={\rm finite}. 
\]
This property holds because in an asymptotically-free theory, $g(\Lambda)$
tends to zero when $\Lambda\rightarrow\infty$. The integral is convergent
around zero, because $\delta_5(g^2)={\cal O}(g^4)$ and $\beta(g)={\cal O}%
(g^3)$.

The full matrix $Z_{ij}$ has a finite limit $C_{ij}(g^{2})$. Using the
Callan-Symanzik equations, we conclude that the operators $J_{5}^{\mu
R}\equiv C^{-1}(g^{2})[J_{5}^{\mu }]$ and ${\cal O}_{i}^{R}\equiv
C_{ij}^{-1}(g^{2})[{\cal O}_{j}]$ have two-point functions of the form 
\[
\langle J_{5}^{\mu R}(x)\,J_{5}^{\nu R}(0)\rangle ={\frac{%
A(g^{2}(1/|x|))\delta _{\mu \nu }+B(g^{2}(1/|x|))\,x^{\mu }x^{\nu }/x^{2}}{%
(x^{2})^{3}}},~~~~\langle {\cal O}_{i}^{R}(x)\,{\cal O}_{j}^{R}(0)\rangle ={%
\frac{A_{ij}(g^{2}(t))}{(x^{2})^{4}}}, 
\]
and admit unambiguous critical limits, as in (\ref{oo2}).

\medskip \medskip

{\large {\bf Acknowledgements}} \vskip .2truecm

I am grateful to G.C. Rossi for discussions on the contents of the appendix,
G. Festuccia for useful remarks, CERN for hospitality during the early stage
of this work and MIT for hospitality during the final stage of this work.

\end{document}